\documentclass[10pt,letterpaper]{article}
\usepackage{amsmath,amssymb}

\usepackage{changepage}

\usepackage{textcomp,marvosym}

\usepackage{cite}

\usepackage{nameref,hyperref}

\usepackage[table]{xcolor}

\usepackage{array}

\usepackage[normalem]{ulem}

\usepackage[utf8]{inputenc}
\usepackage{times}

\usepackage{bbm}
\usepackage{float}

\newcommand{\prox}{\operatorname{prox}}
\newcommand{\sign}{\operatorname{sign}}

\usepackage{amsmath}

\DeclareMathOperator*{\argmin}{arg\,min}

\definecolor{Gray}{gray}{0.85}

\newcolumntype{+}{!{\vrule width 2pt}}

\newlength\savedwidth

\raggedright
\usepackage[aboveskip=1pt,labelfont=bf,labelsep=period,justification=raggedright,singlelinecheck=off]{caption}

\makeatletter
\renewcommand{\@biblabel}[1]{\quad#1.}
\makeatother

\usepackage{lastpage,fancyhdr,graphicx}
\usepackage{epstopdf}
\fancyheadoffset[L]{2.25in}
\fancyfootoffset[L]{2.25in}
\begin{document}
\vspace*{0.2in}

\begin{flushleft}
{\Large
\textbf\newline{Beyond $\ell_1$ sparse coding in V1} % Please use "sentence case" for title and headings (capitalize only the first word in a title (or heading), the first word in a subtitle (or subheading), and any proper nouns).
}
\newline
\\
Ilias Rentzeperis\textsuperscript{1,\textcurrency*},
Luca Calatroni\textsuperscript{2},
Laurent Perrinet\textsuperscript{3},
Dario Prandi\textsuperscript{1}
\\
\bigskip
\textbf{1} Université Paris-Saclay, CNRS, CentraleSupélec, Laboratoire des Signaux et Systèmes, Paris, France
\\
\textbf{2} CNRS, UCA, INRIA, Laboratoire d’Informatique, Signaux et Systèmes de Sophia Antipolis, Sophia Antipolis, France
\\
\textbf{3} Aix Marseille Univ, CNRS, INT, Institut de Neurosciences de la Timone, Marseille, France
\\
\bigskip

\textcurrency Current Address: Instituto de Óptica, CSIC, Madrid, Spain % change symbol to "\textcurrency a" if more than one current address note
% \textcurrency b Insert second current address 
% \textcurrency c Insert third current address

% Use the asterisk to denote corresponding authorship and provide email address in note below.
* ilias.rentzeperis@gmail.com

\end{flushleft}
\section*{Abstract}
Growing evidence indicates that only a sparse subset from a pool of sensory neurons is active for the encoding of visual stimuli at any instant in time. Traditionally, to replicate such biological sparsity, generative models have been using the $\ell_1$ norm as a penalty due to its convexity, which makes it amenable to fast and simple algorithmic solvers. In this work, we use biological vision as a test-bed and show that the soft thresholding operation associated to the use of the $\ell_1$ norm is highly suboptimal compared to other functions suited to approximating $\ell_q$ with $0 \leq q < 1 $ (including recently proposed Continuous Exact relaxations), both in terms of performance and in the production of features that are akin to signatures of the primary visual cortex. We show that $\ell_1$ sparsity produces a denser code or employs a pool with more neurons, i.e. has a higher degree of overcompleteness, in order to maintain the same reconstruction error as the other methods considered. For all the penalty functions tested,  a subset of the neurons develop orientation selectivity similarly to V1 neurons. When their code is sparse enough, the methods also develop receptive fields with varying functionalities, another signature of V1. Compared to other methods, soft thresholding achieves this level of sparsity at the expense of much degraded reconstruction performance, that more likely than not is not acceptable in biological vision. Our results indicate that V1 uses a sparsity inducing regularization that is closer to the $\ell_0$ pseudo-norm rather than to the $\ell_1$ norm.

% Please keep the Author Summary between 150 and 200 words
% Use first person. PLOS ONE authors please skip this step. 
% Author Summary not valid for PLOS ONE submissions.   
\section*{Author summary}
Recordings in the brain indicate that relatively few sensory neurons are active at any instant. This so called sparse coding is considered a hallmark of efficiency in the encoding of natural stimuli by sensory neurons. Computational works have shown that if we add sparse activity as an optimization term in a generative model encoding natural images then the model will learn units with receptive fields similar to the neurons in the primary visual cortex. Traditionally, computational models have used the $\ell_1$ norm as the sparsity term to be minimized, because of its convexity and claims of optimality. Here we show that by using sparsity inducing regularizers that approximate the $\ell_0$ pseudo-norm we get sparser activations for the same quality of encoding. Moreover, for a constrained number of units and a certain level of sparsity we get variability in the functionalities of the receptive fields, a feature of neurons in the primary visual cortex. For this sparsity level , the model using the $\ell_1$ norm has the worst encoding performance. Our study shows that sparsity inducing regularizers approaching the $\ell_0$ norm are more appropriate for modelling biological vision.

% Use "Eq" instead of "Equation" for equation citations.
\section*{Introduction}

Sensory neurons produce a variable range of responses to stimuli, the most frequent one being inactivity~\cite{lettvin_what_1959,burns_uncertain_1968}. To explain this, Horace B. Barlow hypothesized that the task of  sensory neurons is not only to encode in their activity an accurate representation of the outside world, but to do so with the least possible number of active neurons at any time~\cite{barlow_single_1972}. Since then, growing experimental evidence across species and sensory areas has confirmed these claims of sparse activity~\cite{vinje_sparse_2000,perez-orive_oscillations_2002,hromadka_sparse_2008,quiroga_sparse_2008,willmore_sparse_2011}.

Using Barlow's hypothesis as an optimization principle, Olshausen and Field showed that a neural network equipped with a  learning algorithm that is set to reconstruct natural images with sparse activity constraints develops units (atoms) with properties similar to the ones of receptive fields (RFs) of simple cells in the primary visual cortex (V1), i.e.~they are bandpass, oriented and spatially localized~\cite{olshausen_emergence_1996}. The model  proposed by the authors belongs to the family of generative algorithms which represent a stimulus as a linear combination of atoms taken from an overcomplete dictionary, i.e. a set of vectors with more basis vectors than the dimension of the stimuli. In the context of V1, the vectors from the dictionary and their accompanying coefficients correspond to the neurons' RFs and activities, respectively. 

Computationally, overcompleteness comes with a number of advantages: the input can be in a compressed form~\cite{schmid-saugeon_dictionary_2004}, the emerging vectors in the overcomplete dictionary are shiftable, and transformations of the input image such as rotations or translations can be represented by smooth changes in the coefficients~\cite{simoncelli_shiftable_1992}. Experimental findings in the macaque show that overcomplete dictionaries reflect the expansion of inputs in layer 4C$\alpha$ of V1 compared to the lateral geniculate nucleus (LGN) input to it; approximately, $30$ lateral LGN neurons send their axons to a V1 hypercolumn containing about $3000$ excitatory and $1000$ inhibitory neurons~\cite{lund_anatomical_2003,angelucci_contribution_2006,chariker_orientation_2016}.

Sparse approximation aims to find a linear combination of the dictionary vectors that has few nonzero coefficients but also adequately represents the input signal. Ideal sparse approximation requires the minimization of the noncontinuous and nonconvex $\ell_0$ pseudo-norm, which counts the number of nonzero coefficients, combined with some data fitting term. However,  this  problem is NP-hard, as its solution requires an intractable combinatorial search~\cite[p.418]{mallat_wavelet_1998}. Greedy pursuit methods are practical solutions which bear ressemblance to neural spiking processes~\cite{perrinet_role_2010}, yet their efficiency can be improved. In many applications, the $\ell_0$ pseudo-norm has been replaced with its convex relaxation, the $\ell_1$ norm, defined as the sum of the absolute values of the coefficients. The use of the $\ell_1$ relaxation has become widespread in sparse coding, due to its convexity, and since under certain conditions~\cite{candes_stable_2006,candes_restricted_2008} solutions of the $\ell_1$-penalized sparse coding problem coincide with the ones making use of $\ell_0$ regularization. In general, however, $\ell_1$-based models show inferior results in terms of sparsity~\cite{chartrand_exact_2007,candes_enhancing_2008}. 

Over the last decade, advances in optimization theory and in the field of compressed sensing~\cite{Donoho2006} have provided several tools allowing for the replacement of the $\ell_0$ pseudo-norm with tractable functions  approximating it. The use of such approaches provides solutions for many perceptual and behavioral tasks that are in line with the energy constraints in the brain, unlike exact solutions that need perfect prior knowledge and costly computations~\cite{gardner_optimality_2019}.

% \medskip

In this work, we examined different sparse coding algorithms relying on the use of tighter thresholding functions associated to the use of $\ell_q$ penalties, with $ 0 \leq q < 1 $. We found that their solutions induce sparsity to a greater extent compared to the $\ell_1$ method (soft thresholding) while they maintain the same reconstruction of the signal. As a further penalty we used the Continuous Exact $\ell_0$ relaxation (CEL0)~\cite{soubies_continuous_2015} which produced the sparsest codes for small degrees of overcompleteness.

We then analyzed the RFs learned by the resulting sparse coding algorithms and compared them with each other and with the RFs found in the visual cortex of non-human primates. We found that all algorithms yield localized oriented RFs. The tuning width of the RFs showed variability, a characteristic that is also present in V1 cells~\cite{ringach_dl_orientation_2002,gharat_nonlinear_2017}. In accordance with the oblique effect and its representation in the visual cortex~\cite{de_valois_orientation_1982,li_oblique_2003}, we found a preference towards the vertical orientation both in terms of overrepresentation and increased sensitivity of RFs tuned to it.

We show that for a relatively small degree of overcompleteness (approximately $2$ times the size of the input stimulus), the sparse approaches considered produced both sharply and broadly tuned orientation selective RFs, similarly to V1 neurons in the cat and the macaque~\cite{ringach_spatial_2002,gharat_nonlinear_2017}. In terms of performance, when keeping sparsity constant for all methods, we found that soft thresholding requires $10$ times more units (a $20\times$ degree of overcompleteness) to reconstruct the input image patch as well as CEL0. The other methods considered, relying, e.g., on  $\ell_{1/2}$ minimization~\cite{xu__2012} and  on the hard thresholding algorithm~\cite{blumensath_iterative_2008} are inferior to CEL0 in terms of reconstruction performance but still superior to soft thresholding, and show more robust convergence to a solution for a dictionary with larger degrees of overcompleteness compared to CEL0. 

By definition, $\ell_1$ regularization employed by soft thresholding limits average neural activity (and subsequently metabolic energy consumption) rather than the number of neurons~\cite{olshausen_emergence_1996,rehn_network_2007}. Here, we show that the regularizers pushing for less units (not smaller activities) develop RFs with variability in their response patterns, a feature which is present in V1~\cite{ringach_spatial_2002,gharat_nonlinear_2017,ladret_dynamical_2022}. In line with our results, a recent study on mice showed indeed that natural images could be decoded from a very small number of highly active V1 neurons, and that diverse RFs ensure both reliable and sparse representations~\cite{yoshida_natural_2020}.

Neural activation and learning based on $\ell_1$ minimization could in theory accommodate the variability in V1 since there is a $100$ fold size expansion from LGN to V1~\cite{lund_anatomical_2003,angelucci_contribution_2006,chariker_orientation_2016}. But this $100\times$ degree of overcompleteness is misleading since V1 neurons are divided into many different pools with different objectives and follow different visual streams~\cite{zeki_functional_1978,hubel_segregation_1987,livingstone_segregation_1988}, even though there may be some overlap in their functionalities~\cite{rentzeperis_distributed_2014}. So with the constraint that the expansion of V1 neurons compared to LGN is divided for different image reconstruction tasks, the observed optimality of brain processes suggests that neural activation is regularized in such a way that it produces the variability of RFs found in the visual cortex with the smallest degree of overcompleteness. This in turn indicates that V1 uses heuristics that are closer in approximating $\ell_0$ penalties rather than $\ell_1$.

\section*{Materials and methods}\label{sec:methods}
\subsection*{Image dataset and preprocessing}  \label{sec:dataset}
For our tests, we used a selection of $137$ natural images from the van Hateren's database~\cite{van_hateren_independent_1998}. These images did not contain artificially created structures neither significant blur, similarly as in~\cite{graham_can_2006}. We performed the same preprocessing stage described in~\cite{olshausen_sparse_1997}: first, we rescaled all images separately between zero and one, then we normalized them by subtracting and dividing each pixel value by the image mean and standard deviation respectively. The resulting zero mean, unit variance images were then passed through a whitening filter in order to emulate the response of retinal ganglion cells. The images were finally rescaled so that they have a variance of $0.1$. This value serves as a baseline error, i.e. the mean square error (MSE) of a preprocessed image with an image with only zero pixel values (produced when all the coefficients of a neural code are zero).  \nameref{S1_Fig} shows examples of raw and preprocessed images. 
\subsection*{Sparse coding generative models} \label{sec:generative_models}
\subsubsection*{Model setup and cost function}
According to the linear generative model of Olshausen and Field~\cite{olshausen_emergence_1996}, an image   $I\in\mathbb{R}^M$ is described as a linear combination of vectors $(\phi_i)_{i=1}^N$ with $\phi_i\in\mathbb{R}^M$ for all $i$. The vectors $(\phi_i)_{i=1}^N$ are stored column-wise in a matrix  $\Phi\in\mathbb{R}^{M\times N}$. The scalar coefficients of such linear combination are collected in a vector $r\in\mathbb{R}^N$ and an additive white Gaussian noise component $\nu\in\mathbb{R}^M$ with $\nu\sim \mathcal{N}(0,\sigma^2 \text{Id})$ is added to model perturbations and uncertainty:
\begin{equation} \label{eqGenModel}
I = \Phi r + \nu = \sum_{i=1}^N r_i \phi_i + \nu.
\end{equation}
%Here, $r_i$ is the $i^{th}$ coefficient in $r$ and $\phi_i$ the $i^{th}$ column of $\Phi$. 

We consider features (columns) $\Phi$ to form an overcomplete dictionary, i.e.~$N\gg M$. Consequently, the inverse problem of finding $r$ given $I$ in \eqref{eqGenModel}
% when $\Phi$ is known,
becomes ill-posed since $r$ may have an infinite number of possible solutions. % To {\color{red}reduce the number of solutions} \dario{impose a sort of  well-posedness} 
To impose well-posedness,
Olshausen and Field~\cite{olshausen_emergence_1996} considered a sparse regularisation approach,  defined in terms of the energy function:
\begin{equation} \label{eqEnergy}
E(r,\Phi): = {\frac{1}{2} \|I-\Phi r \| ^2_2+ \lambda \sum_{i=1}^N   c(r_i)}.
\end{equation}
While the first term in \eqref{eqEnergy} pushes towards the preservation of stimulus information, the second term  acts as regularization imposing a penalty on activity with the relative weight of the two competing tasks being controlled by a parameter $\lambda>0$. The regularization term $C(r):= \sum_{i=1}^N c(r_i)$ is a sparsity-promoting penalty that ideally encourages the number of active units to be as few as possible. For that, one would like to choose as $c(\cdot)$ the so-called  $\ell_0$ pseudo-norm of $z$ which costs $1$ whenever $z\neq 0$ and $0$ otherwise: $c(z)=||z||_0$, with $z\in \mathbb{R}$. However, as shown rigorously in several mathematical works (e.g.,~\cite{Natarajan1995}) such choice makes the problem  of minimising $E$ in \eqref{eqEnergy} NP-hard. A standard strategy, used in several sparse coding approaches, relies on the use of the convex and continuous $\ell_1$
norm as a relaxation, i.e., $c(z)=|z|$ for $z\in\mathbb{R}$. Under suitable conditions on the matrix $\Phi$, such choice guarantees indeed that the solution computed is equivalent to the one corresponding to the $\ell_0$ pseudo-norm. The use of the $\ell_1$ norm is in fact  established in the field of compressed sensing and sparse signal/image processing~\cite{Candes2008}. 

For general choices of $c(\cdot)$, the problem of finding both optimal sparse codes $r^*$ (coding step) and feature vectors $\Phi^*$  for the given input stimulus $I$ (learning step) can be formulated as the problem of minimizing the energy function $E$ with respect to both $r$ and $\Phi$, i.e:
\begin{equation} \label{eqArgMin}
(r^*,\Phi^*) \in  \underset{r \in \mathbb{R}^N,\Phi\in\mathbb{R}^{M\times N} }{\arg\min} \, E(r,\Phi).
\end{equation}
In the following, we use alternating minimization (see, e.g., \cite{Wang2008}) to solve the problem above. Below, we thus make precise the general approach for solving the coding and learning steps. Subsequently, we consider few cost functions promoting sparsity in different ways.

\subsubsection*{Coding step}%}
Our objective is to minimize the composite function $E(r,\Phi)$ in \eqref{eqEnergy} which is defined as the sum
\begin{equation} \label{eqCompFunction} 
E(r,\Phi) = f(r,\Phi) + \lambda  C(r),
\end{equation}
%iterative thresholding functions alternate between a gradient step reducing the convex objective function, $f(x)$, and a thresholding step that is typically derived by $g(x)$. 
with $f:\mathbb{R}^N \times \mathbb{R}^{M\times N}\to\mathbb{R}_+$ being convex and differentiable with $L$-Lipschitz gradient w.r.t.~both variables and $C:\mathbb{R}^N\to\mathbb{R}_+$ being convex, proper and lower semi-continuous, but, generally, non-smooth. 
As far as the coding step is concerned, we then need an algorithm solving the structured nonsmooth optimisation problem \eqref{eqArgMin} w.r.t.~$r$.
A standard strategy for this is to use the proximal gradient algorithm (see~\cite{Beck2017} for a review). For a given step-size $\mu\in(0,1/L]$, $x_0\in\mathbb{R}^N$, $\bar{\Phi}\in\mathbb{R}^{M\times N}$ and $t\geq 0$, 
such algorithm consists in the alternative application of the two steps:
\begin{itemize}
\item Gradient step: $ r_{t+1} = r_t - \mu \nabla f(r_t,\bar{\Phi})$;
\item Proximal step: $r_{t+1} = \prox_{\mu, \lambda C}(r_{t+1}) $,
\end{itemize}
where the proximal operator associated to the function $\lambda C(\cdot)$ and depending on the step-size parameter $\mu$ is defined by:
\begin{align}   \label{eq:prox}
\prox_{ \mu, \lambda C}(z) & = \argmin_{y\in\mathbb{R}^N}~ \lambda C(y) + \frac{1}{2\mu}\|y-z\|^2 \\
& = \argmin_{y\in\mathbb{R}^N}~ C(y) + \frac{1}{2\lambda\mu}\|y-z\|^2   \notag \\
& = \prox_{ 1, \mu\lambda C}(z), \quad z\in\mathbb{R}^N.  \notag
\end{align}
It is common to denote by $T_{ \mu \lambda}(z)$ the thresholding operation corresponding to $\prox_{\mu, \lambda C}(z)$, which sets to zero components of $z$ which are too large depending on a certain thresholding rule defined in terms of the choice of $C$ and the thresholding parameter $\mu$, see Section~\ref{sec:thresh} for some examples. 
%The relation between the two functions can be made more precise using tools from convex analysis such as sub-differentials and resolvent operators, see, e.g.,~\cite{ChambollePock2016} for more details.

%In the gradient step $\mu$ is a step size, 
%and in the sparsity step $T_{\mu}(\cdot) $ is a thresholding operator that performs a sparse approximation of the coefficients produced from the gradient step. If $g(x)$ is convex, but eventually noncontinuous, the thresholding operator is equal to the proximal one, i.e.~the sparsity step becomes $ x_{t+1} = \prox_{\mu g}(x^*_{t+1})$.

Note that during coding, the vectors in $\bar{\Phi}$ are kept fixed, so that the algorithm seeks the optimal activations for the given input image patches. 

%We follow the procedure described above to derive the coding step from \eqref{eqEnergy}. The iterative procedure can be thus specified as follows:
%\begin{enumerate}
%    \item $r_0$ is initialized;
%    \item $r^*_{t+1}=r_{t}+\mu \Phi^T(I-\Phi r_t)$;
%    \item $r_{t+1} = T_{\mu\lambda}(r^*_{t+1})$;
%    \item Repeat steps 2 and 3 until convergence.
%\end{enumerate}
%In the scheme above, the form of the thresholding operator $T_{\mu\lambda}(\cdot)$ depends on the particular choice of the function $\lambda \sum C(\cdot)$. 

\subsubsection*{Learning step} \label{sec:learning} 
During learning, the matrix $\Phi\in\mathbb{R}^{M\times N}$ is updated so that it is optimal in reconstructing the input $I$ as accurately as possible. The learning step is thus obtained by minimizing \eqref{eqEnergy} w.r.t.~$\Phi$, by considering gradient-type iterations. This step is easier since $\Phi$ appears only in the smooth data fit term and not in the cost term.
%is not in the cost function $C(\cdot)$ {\color{red}(does not depend on $g(x)$ in equation \eqref{eqCompFunction})} \dario{(i.e.~$g(x)$ in equation \eqref{eqCompFunction} does not depend on $\Phi$)} 
%and we take the gradient only on the convex first part of equation \eqref{eqEnergy}. 
Learning is then obtained for all $t\geq 0$ via the iterative procedure:
\begin{equation} \label{eqLearning}
\Phi_{t+1} = \Phi_{t} + \eta  r (I - \Phi_{t} r),
\end{equation}
where $ \eta>0$ is the learning rate whose size has to be small enough to guarantee convergence. Note that, although such update of $\Phi$ does not depend explicitly on the particular choice of the cost function considered, it depends nevertheless on the current estimate of the coefficients $r$ which, in turn, depend on the particular choice of $C$ and, consequently, $T_{\mu\lambda}$. During each learning step, we impose the norms of the current iterate $\Phi_{k}$ to be equal to $1$, though other normalization mechanisms could be explored~\cite{perrinet_adaptive_2019}.

\subsection*{Thresholding operators}  \label{sec:thresh}

\begin{figure}[h!]
  \centering
      \includegraphics[width=\textwidth]{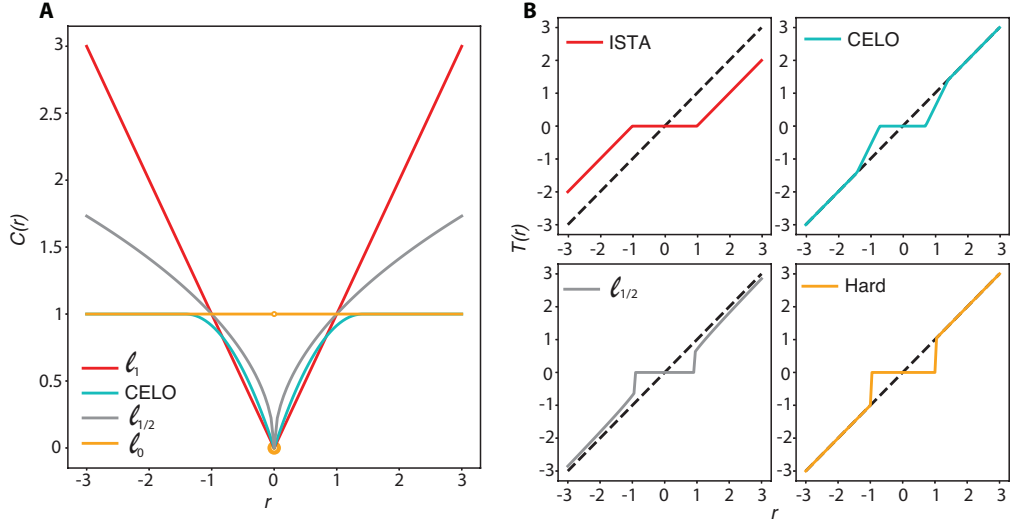}
  \caption{{\bf Sparsity-promoting cost functions $c$ considered and their corresponding thresholding operators.} A: Plot of 1D cost functions $c:\mathbb{R}\to \mathbb{R}_+$ considered. B: Corresponding thresholding operators. We denoted by ISTA the operator $S_\theta(z)$ being the thresholding operator associated to the use of $\ell_1$ and by Hard the operator $H_{\theta}$ associated to the iterative hard thresholding algorithm.}
  \label{fig:Fig1}
\end{figure}

For different choices of the component-wise cost functions $c:\mathbb{R}\to\mathbb{R}_+$, different thresholding rules are derived. We consider below some particular choices of $c$, plotting them in Fig~\ref{fig:Fig1}A for comparison. For each choice, we then report the corresponding explicit thresholding operator which sets to zero coefficients with small magnitudes, see Fig~\ref{fig:Fig1}B for an illustration. As a technical note, we remark that in definition \eqref{eq:prox} we assumed the function $C$ is convex, so that the minimiser of the functional is uniquely defined due to the strong convexity of the composite function. A large class of the cost functions considered below, however, are not convex, hence definition  \eqref{eq:prox}  still holds, but with a $\in$ sign in place of the equality one, since the set of minimizers may not be a singleton.
%\dario{In the following we present the different thresholding operators considered in our analysis.}

\subsubsection*{The iterative soft-thresholding algorithm (ISTA)}

For $c(r_i)=|r_i|$ for all $i=1\ldots,N$ sparsity in \eqref{eqEnergy} is obtained by considering as regulariser the function $C(r)=\lambda\|r\|_1$ which, from an algorithmic viewpoint, corresponds to the Iterative Soft-Thresholding Algorithm (ISTA)
as an algorithmic solver~\cite{daubechies_iterative_2004,blumensath_iterative_2008}. Thanks to the separability of the $\ell_1$ norm, the proximal operator can be computed component-wise~\cite{parikh_proximal_2014}. Setting $\theta:=\mu\lambda>0$ and  $S_\theta(\cdot):=T_{\theta}(\cdot)$, it holds that for all $i=1,\ldots,N$:
\begin{equation} \label{softEqu}
S_\theta(r^*_i)= 
\begin{cases} 
r^*_i-\theta & (r^*_i>\theta)\\ 
0 & (-\theta\leq y\leq\theta)\\ 
 r^*_i+\theta & (r^*_i<-\theta).
\end{cases}
\end{equation}
Such operation is typically known in literature under the name of soft-thresholding operator due to its continuity outside the vanishing thresholding region.

\subsubsection*{The iterative half thresholding algorithm }%}

To favour more sparsity than the $\ell_1$ norm, a natural improvement consists in using the $\ell_q$ $(0<q<1)$ pseudo-norm, i.e.~setting $c(r_i)=|r_i|^q$. However, such choice makes the optimization problem \eqref{eqEnergy} nonsmooth and nonconvex and, in general, prevents from using fast optimisation solvers. One exception to this is the case when $\ell_{1/2}$ regularization is used. Xu and colleagues showed in~\cite{xu__2012} that an iterative half-thresholding algorithm can solve the problem of minimising the $\ell_{1/2}$ pseudo-norm with an $\ell_2$ data fit term with the algorithm converging to a local minimizer in linear time~\cite{zeng__2014}. Using analogous notation $\theta=\lambda\mu$ as before and denoting by $ \Xi_{\theta, 1/2}(\cdot) $ the thresholding operator $T_{\theta}$,  thresholding can still be performed component-wise as follows:
\begin{equation} \label{eq5}
  \Xi_{\theta, 1/2}(r^*_i) =
    \begin{cases}
      f_{\theta, 1/2}(r^*_i),& |r^*_i|>\sqrt[3]{54}~\theta^{2/3}\\
      0, & \text{otherwise},
    \end{cases}       
\end{equation}
where:
\begin{equation} \label{eq6}
f_{\theta, 1/2}(r^*_i) = 
{\frac{2}{3}r^*_i\left(1 + \cos\left(\frac{2\pi}{3} - \frac{2}{3}\psi_{\theta}(r^*_i)\right)\right)}
\end{equation}
and
\begin{equation} \label{eq7}
\psi_{\theta}(r^*_i)) = \arccos\left(\frac{\theta}{8}\left(\frac{|r^*_i|}{3}\right)^{-3/2}\right).
\end{equation}
Despite its complex form, the thresholding function \eqref{eq5} is still explicit, hence its computation is very efficient.

\subsubsection*{Iterative hard thresholding}%}
The iterative hard thresholding algorithm has been introduced firstly in~\cite{blumensath_iterative_2008} to overcome the NP-hardness associated to the minimization of the ideal problem:
\begin{equation} \label{eq9}
\argmin_{r\in\mathbb{R}^N}~ E_{\ell_0}(r) =  {\|I- \bar{\Phi} r \| ^2_2+ \lambda\|r \|_0 },
\end{equation}
for $\bar{\Phi}\in\mathbb{R}^{M\times N}$.
The idea consists in considering the following surrogate function defined for $r,z\in\mathbb{R}^N$ as:
\begin{equation} \label{eq8}
E_{\ell_0}^{S}(r,z): = 
{\|I-\bar{\Phi} r \| ^2_2+ \lambda\|r \|_0 - \|\bar{\Phi} r - \bar{\Phi} z\| ^2_2+
\|r - z \| ^2_2},
\end{equation}
for which there trivially holds $ E_{\ell_0}(r) = E_{\ell_0}^{S}(r,r) $ for all $r\in\mathbb{R}^N$. Minimizing  \eqref{eq8} with respect to $r$ can thus be seen as a strategy to minimize \eqref{eq9}, as desired. The derived thresholding operator is here denoted by $H_{\theta}(\cdot)$ and performs element-wise hard thresholding following the rule:
\begin{equation} \label{eqGenModel1}
  H_{\theta}(r^*_i) =
    \begin{cases}
      0,& |r^*_i| \leq  \sqrt{\theta}\\
      r^*_i, & \text{otherwise}.
    \end{cases} \end{equation}

\subsubsection*{A continuous exact $\ell_0$ penalty (CEL0)}%}
As a further sparsity-promoting regularization, we consider the non-convex Continuous Exact relaxation of the $\ell_0$ norm (CEL0), thoroughly studied, e.g., in~\cite{soubies_continuous_2015}. Such choice can be thought of (as it is rigorously proved in~\cite{soubies_unified_2017}) as the inferior limit of the class of all continuous and non-convex regularizations of the $\ell_0$ pseudo-norm with the interesting additional properties of preserving the global minimizers of the ideal $\ell_2$-$\ell_0$ minimization problem one would need to solve, while reducing the number of the local ones. For all $i=1,\ldots,N$ and parameter $\lambda>0$ such choice corresponds to considering as cost functional:
\begin{equation} \label{eqGenModel2}
C_{CEL0}(r) := \sum_{i =1}^N c(\| \phi_i\|, \lambda,r_i) = 
\sum_{i=1}^N \left( \lambda - \frac{\| \phi_i \|^2}{2}\left(|r_i|-\frac{\sqrt{2 \lambda}}{\| \phi_i \|}\right)^{2}_+ \right),
\end{equation}
where $\phi_i$ is the $i$-th column extracted from the matrix $\Phi$, $\| \phi_i\|$ denotes its $\ell_2$ norm and, for all $z\in\mathbb{R}$, the notation $(z)_+$ denotes the positive part of $z$, i.e. $(z)_+ = \max(0,z)$. The corresponding CEL0 thesholding operator is defined by:
%\begin{equation} \label{eqGenModel3}
%  \rho_+ =
%    \begin{cases}
%      \rho, & \rho > 0\\
%      0, & \text{otherwise}
%    \end{cases} \end{equation}
\begin{equation} \label{threshCel0}
 \Theta^{\text{CEL0}}_{\mu,\lambda}(r_i^*)= 
    \begin{cases}
        \displaystyle\sign(r_i^*)\min \left\{|r_i^*|,\frac{(|r_i^*|- \sqrt{2\lambda}\mu \| \phi_i\|)_+}{1-\|\phi_i\|^2\mu}\right\}, &  \|\phi_i\|^2\mu < 1 \\[2em]
        r_i^*\mathbbm{1}_{|r_i^*|>\sqrt{2\mu\lambda} }(r_i^*)+\{0,r_i^*\}\mathbbm{1}_{|r_i^*| = \sqrt{2\mu\lambda}}(r_i^*), & \|\phi_i\|^2\mu \geq 1,
    \end{cases}
\end{equation}
where, note, $\mu$ and $\lambda$ are here decoupled as the thresholding parameter is not their parameter anymore but depends on $\mu$ only and, component-wise, by the norm of the column $\phi_i$, $i=1,\ldots,N$ of $\Phi$. While the operation of computing the quantities $\| \phi_i\|$ can be in principle costly from a computational viewpoint, we remark that by construction, in our application $\Phi$ has unit-norm columns, hence such computation is in fact not required.

Our objective for the following sections consists in comparing the sparse coding obtained by the different choices of the cost functions and thresholding functions above. In particular, as the coding step affects the learning step, our study focuses on how the choice of the sparsity-promoting penalty (related to the capability of approximating the $\ell_0$ pseudo-norm) affects the estimation of the receptive fields $\phi_i$ at convergence. In most of our simulations, the degree of overcompleteness of the dictionary is $2\times$, i.e. the number of atoms ($N$) is approximately twice as much as the size ($M$) of the image patch the algorithm aims to reconstruct ($500$ atoms for a $16\times 16$ patch vectorized into a $256$ element vector). Previously, it has been shown for ISTA that a highly overcomplete dictionary, or a very sparse code ---for a dictionary with fixed atoms--- produces RFs with different functionalities~\cite{olshausen_learning_2009,olshausen_highly_2013}. In the following tests, we  constrain the number of atoms in the dictionary ($2\times$ degree of overcompleteness), and test whether the different algorithms considered produce adequate sparse codes generating receptive fields with different functionalities while at the same time having an acceptable reconstruction performance.

\subsection*{A measure for orientation selectivity:  circular variance} \label{sec:circ_var}

To probe the orientation selectivity of each $(\phi_i)_{i=1}^N$ vector (unit) estimated by the different models above as well as to compare them with each other and with experimental data in V1, we used the circular variance measure ($V\in[0,1]$)~\cite{mardia_statistics_1975,batschelet_circular_1981,ringach_dl_orientation_2002}. A unit with a zero circular variance responds only to a particular orientation; a unit with a circular variance of one responds to all orientations equally. Values in between show some selectivity, with the ones closer to one showing a broader orientation selectivity compared to the ones closer to zero.   

The circular variance of a vector  $\phi$ is defined as $V := 1 - |R|$ where $R$ is:
\begin{equation} \label{CircVariance}
	R = \frac{\sum_k  \alpha_{k} e^{i2\theta_{k}}}{\sum_k  \alpha_{k}},
\end{equation}
with $\alpha_{k}$ being the response of the unit at the orientation $\theta_{k}$ ($\theta_{k}$ goes from 0 to $\pi$ in $k = 36$ equidistant steps). A plot of $\alpha_{k}$ as a function of $\theta_{k}$ for a unit corresponds to its orientation tuning curve.

Below we explain how we get the $\alpha_{k}$ values  for all $\theta_{k}$ orientations for a unit~$\phi_{i}$. We first find the spatial frequency for which the unit responds the most by taking the inner product of $\phi_{i}$ with a bank of sinusoidal gratings of different spatial frequencies, orientations, and phases. The optimal spatial frequency is the spatial frequency of the grating giving the highest value. We then proceed by testing gratings with this spatial frequency. We take the inner product of $\phi_{i}$ with the set of gratings with this spatial frequency, orientation $\theta_{j}$ and all the different phases. We gather from that the highest value, yielding $\alpha_{j}$. We repeat this process for all the orientations $\theta_{k}$. We then proceed in estimating the circular variance for unit $\phi_{i}$ from equation \eqref{CircVariance}.

%

% Results and Discussion can be combined.
\section*{Results}

\subsection*{Sparsity of codes}

\begin{figure}[h!]
  \centering
      \includegraphics[width=\textwidth]{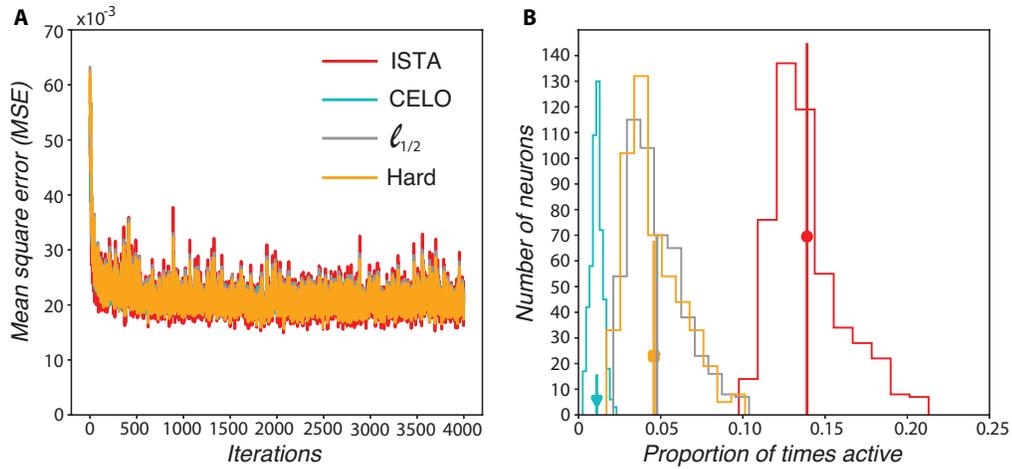}
  \caption{{\bf For the same reconstruction error, CEL0, $\ell_{1/2}$, and hard thresholding produce sparser codes compared to ISTA.} A: MSE between the reconstructed and the actual image as a function of iterations for the different thresholding algorithms. B: Distribution of activity of the units for the image stimuli presented. The middle of the vertical lines represent the mean number of active neurons per image patch, their length the standard deviation}
  \label{fig:Fig2}
\end{figure}

We first compared the sparsity of the codes produced by the different penalty functions above, each containing $500$ units ($\sim$ $2\times$ degrees of overcompleteness). To make a fair comparison, we adjusted the methods' parameters $\lambda$ and $\mu$ so that they all produce reconstructed images with the same MSE (the values of the parameters are shown in \nameref{tableParams500}). We run the algorithms for $4000$ batches, with each batch containing  $250$ image patches. In all cases, the MSE for the last $500$ batches is about $0.021$ (see Fig ~\ref{fig:Fig2}A). As expected from preprocessing, the baseline error, i.e.~the MSE corresponding to the image where all activations are zero is $0.1$. We found that for the same MSE, CEL0, $\ell_{1/2}$, and hard thresholding algorithms produce sparser codes compared to ISTA, with $\ell_{1/2}$, and hard thresholding having very similar activity distributions and CEL0 being the approach corresponding to the sparsest solutions (Fig ~\ref{fig:Fig2}B). Unless otherwise stated, the same $\lambda$ and $\mu$ parameters were employed in subsequent experiments.

The vectors in $(\phi_i)_{i=1}^N$ are updated (learning step) at each iteration on a batch of image patches. As discussed in Section~\ref{sec:learning}, the different thresholding algorithms produce vectors $\phi_i$ showing several differences. We thus asked whether the different penalties considered have similar coding performances -- in terms of sparsity and reconstruction error -- when the vectors $\phi_i$ are not learned during coding. We probed this question by using throughout the coding steps a fixed  dictionary $\Phi$ (we considered, in particular all dictionaries output at convergence of all four algorithms). We observed a sort of invariance property with respect to the dictionary used: all thresholding algorithms show similar reconstruction error and distributions of activity independently of the dictionary used in coding (\nameref{S2_Fig}).
 
\begin{figure}[h!]
  \centering
      \includegraphics[width=\textwidth]{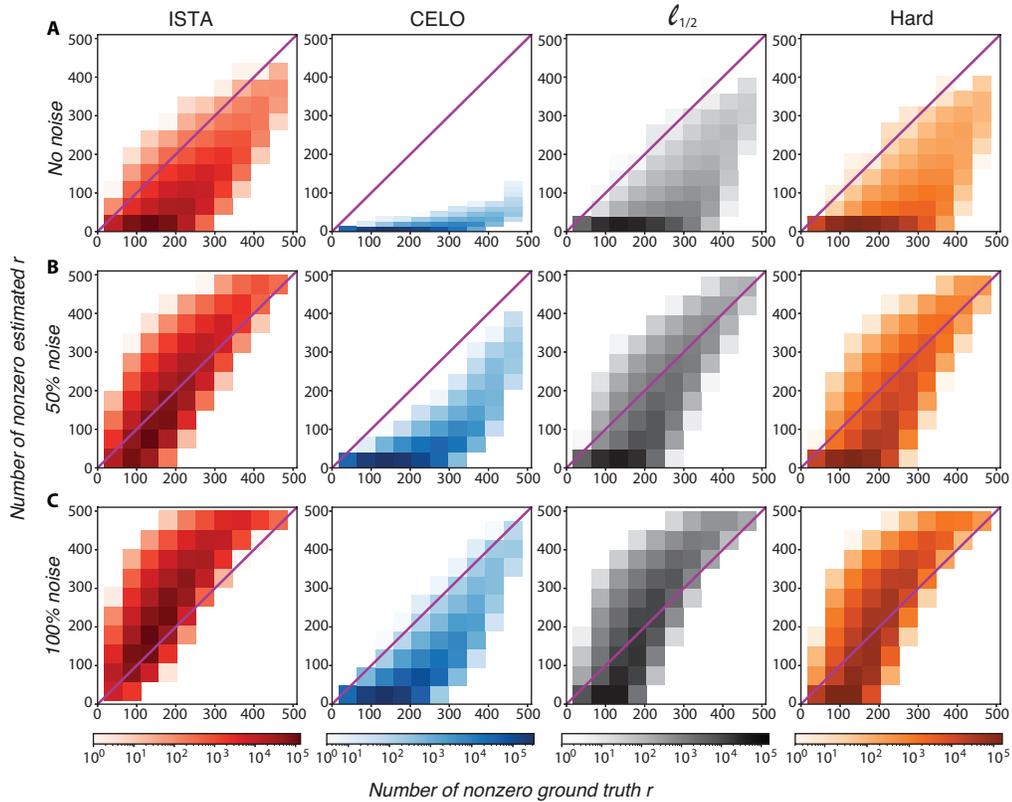}
  \caption{{\bf For moderate noise levels (50\%) all methods produce on average sparser codes compared to the ground truth; for extreme noise levels (100\%) only CEL0 produces sparser codes.} The figures show the number of nonzero elements of the reconstructed $r^*$ as a function of the actual nonzero elements in the ground truth vectors $\bar{r}$ for the different methods (columns) and noise levels (rows). Points above the diagonal indicate examples where the number of estimated nonzero elements are greater than the actual ones. A: No noise B: $50\%$ noise C: $100\%$ noise. Noise is white additive Gaussian scaled by the $\ell_{1}$ norm of the of the $\bar{r}$'s times the fraction of noise.}
  \label{fig:Fig3}
\end{figure}

In general, the sparse codes from each method provide an estimation of the underlying sources of the original images. To compare the sources generated by the methods (estimated $r^*$ sequences) with the actual ones, we created images from known sources (ground truth $\bar{r}$ sequences). More specifically, we created ground truth images by multiplying each of the ground truth vectors $\bar{r}$ with a fixed dictionary $\Phi_{ISTA}$ and adding different noise levels. The vectors $\bar{r}$ were obtained by permuting a subset of vectors $r$, pre-computed by running ISTA and all producing input images with a mean square pixel value over $0.15$, so that the zero solution is less probable to be optimal for $r^*$. For each method the $\lambda$ parameters were the same as in the previous experiments. We found that in the absence of noise, ISTA found codes that are sparser than the ground truth ones for most of the input image patches, with $\ell_{1/2}$ and hard thresholding producing even sparser solutions, and CEL0 producing the sparsest codes (Fig ~\ref{fig:Fig3}A). As we added noise to the images (while keeping the $\lambda$ parameter fixed), such ranking between algorithms in terms of sparsity did not change, but the number of nonzero activations they produced increased, with CEL0, $\ell_{1/2}$ and hard thresholding producing on average sparser codes compared to the ground truth for moderate levels of noise (50\% noise), and with only CEL0 producing sparser codes for very noisy inputs (100\% noise)  (Fig ~\ref{fig:Fig3}B and C).

\subsection*{Learned dictionaries and their connections to V1} % \label{sec:learning}

\subsubsection*{Orientation selectivity of RFs}

\begin{figure}[h!]
  \centering
      \includegraphics[width=\textwidth]{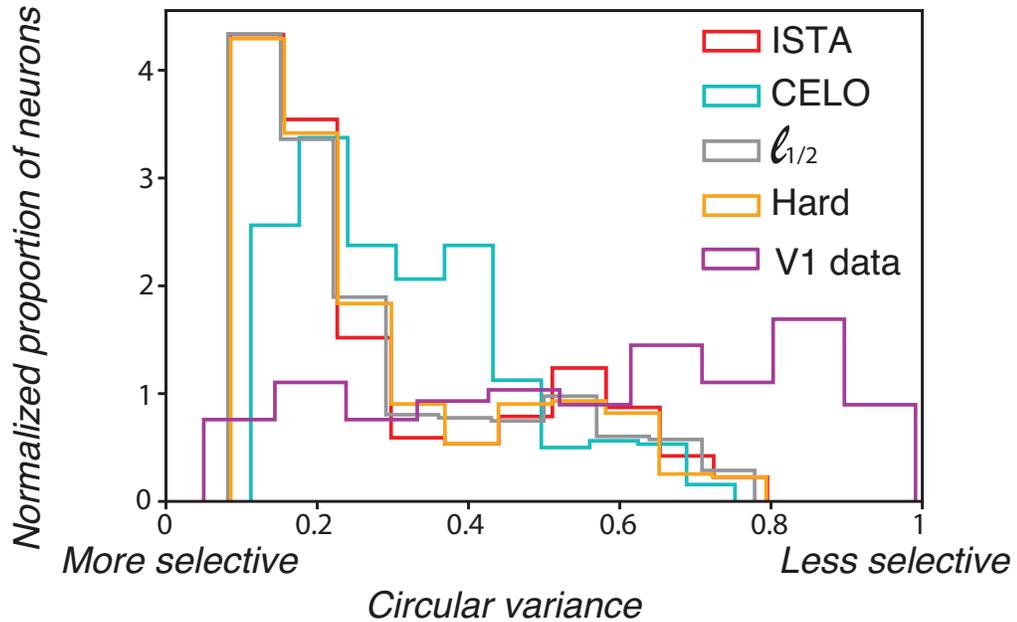}
  \caption{{\bf CEL0 produces vectors $\phi_i$ with a more uniform distribution of circular variance values compared to the other thresholding methods.} Distribution of circular variance values for the $\phi_i$ learned by the different thresholding algorithms (the area in all cases was normalized to sum to 1). We also include V1 experimental data from~\cite{ringach_dl_orientation_2002}}
  \label{fig:Fig4}
\end{figure}

Experimental evidence indicates that neurons in V1 show great variability in their orientation selectivity: some neurons respond to a narrow band around a particular orientation, but most of them are responsive to a broader spectrum of orientations~\cite{ringach_dl_orientation_2002,gharat_nonlinear_2017}. To examine the orientation selectivity of the  vectors $\phi_i^*$  produced by each thresholding algorithm and draw comparisons with experimental data, we used the circular variance measure (\cite{mardia_statistics_1975,batschelet_circular_1981,ringach_dl_orientation_2002} defined in Section~\ref{sec:circ_var}). We found that the use of the $\ell_1$ as well as the $\ell_{1/2}$ penalty, and hard thresholding algorithms, have all a similar distribution of circular variances, while CEL0 has a smaller peak at low circular variance values and a more gradual decrease for higher values (Fig ~\ref{fig:Fig4}) resulting overall in a more `spread' distribution. This is in agreement with experimental data: populations of V1 neurons in the macaques do not show a sharp peak for low circular variance values but rather a more uniform distribution across small and large values (Fig ~\ref{fig:Fig4}; data taken from~\cite{ringach_dl_orientation_2002}).

\begin{figure}[h!]
  \centering
      \includegraphics[width=\textwidth]{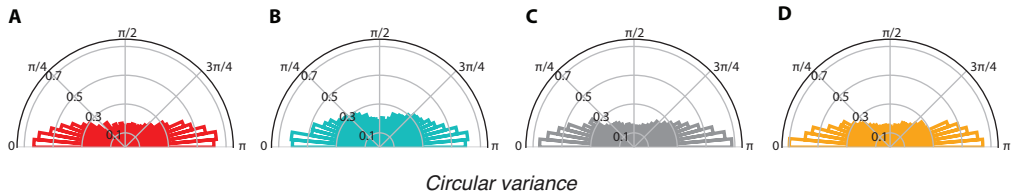}
  \caption{{\bf The vectors $\phi_i$ with the smallest circular variance values (narrower tuning width) respond maximally in the vertical orientation.} Polar plots of the mean circular variance of $\phi$ vectors binned according to their prefered orientation for A: ISTA B: CEL0 C: $\ell_{1/2}$ and D: Hard thresholding.}
  \label{fig:Fig5}
\end{figure}

\begin{figure}[h!]
  \centering
      \includegraphics[width=\textwidth]{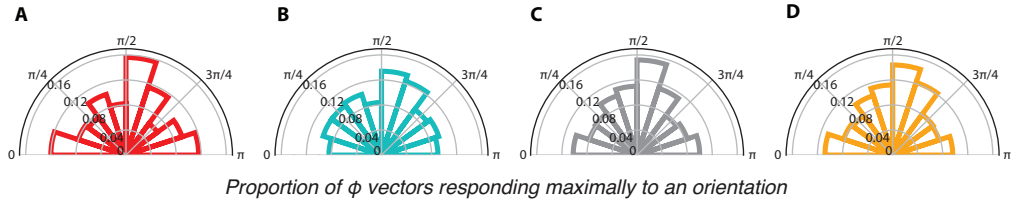}
  \caption{{\bf The largest proportion of vectors $\phi_i$ responding maximally to the vertical orientation.} Polar plots of the proportion of $\phi$ vectors responding maximally to an orientation for A: ISTA B: CEL0 C: $\ell_{1/2}$ and D: Hard thresholding.}
  \label{fig:Fig6}
\end{figure}

Perceptually, visual stimuli are better resolved when they are presented in the cardinal orientations-either horizontal or vertical-as opposed to oblique ones~\cite{appelle_perception_1972}. This behavioural oblique effect has been suggested to emerge in part due to the over-representation of simple cells in V1 that respond to cardinal orientations as shown by single unit recordings~\cite{de_valois_orientation_1982,li_oblique_2003}, optical imaging~\cite{chapman_overrepresentation_1998} and fMRI~\cite{furmanski_oblique_2000}. Moreover, single unit recordings indicate that cardinal orientations have narrower orientation tuning curves~\cite{li_oblique_2003}. We found that for all algorithms, the vectors $\phi_i$ responding to the vertical orientation ($\pi/2$) had the narrowest orientation tuning curves as indicated by their low circular variance value (Fig ~\ref{fig:Fig5}). The proportion of vectors $\phi_i$ responding maximally to the vertical orientation is, moreover, the highest compared to all the other orientations (Fig ~\ref{fig:Fig6}). Both results agree with the experimental results. 

\subsubsection*{Sparsity-induced variability of RFs}
\begin{figure}[h!]
  \centering
      \includegraphics[width=\textwidth]{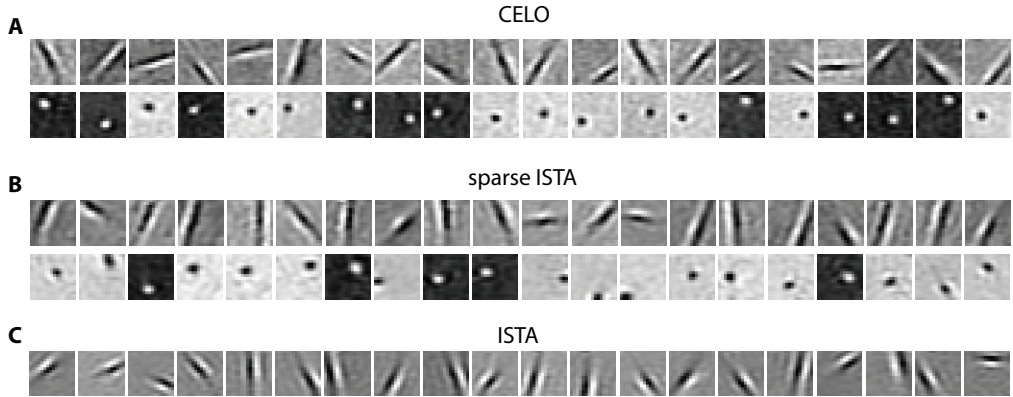}
  \caption{{\bf Only sufficiently sparse algorithms  produce circular RFs.} A: RFs found using CEL0. First row: examples of RFs that are bandpass and oriented. Second row: examples of RFs that show circular selectivity. B: RFs found by ISTA after tuning of parameter $\lambda$ to be as sparse as CEL0. First and second row show examples of the same kind of RFs as in (A). C: Examples of bandpass oriented RFs produced by ISTA  in our initial setup.}
  \label{fig:Fig7}
\end{figure}

More striking differences can be observed after a visual inspection of the corresponding RFs (\nameref{S3_Fig}, \nameref{S4_Fig}, \nameref{S5_Fig}, and \nameref{S6_Fig}). In particular, we see that CEL0 not only produces oriented RFs like the other algorithms but also circular RFs not generally found when using the other algorithms (Fig ~\ref{fig:Fig7}A). For ease of presentation, we focus here on a comparison between CEL0 and ISTA.

\begin{figure}[h!]
  \centering
      \includegraphics[width=\textwidth]{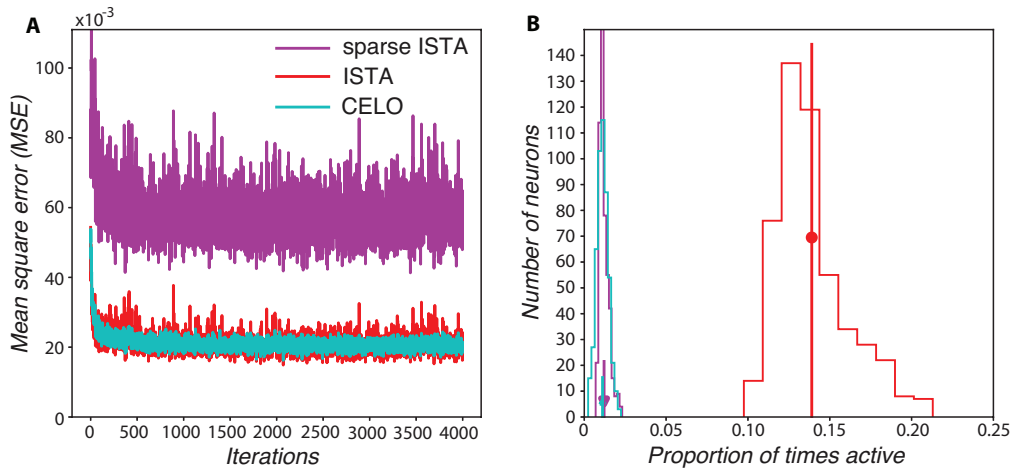}
  \caption{{\bf When ISTA matches in sparsity CEL0 its performance in terms of MSE becomes much worse.} A: MSE between the reconstructed and the actual image as a function of iterations. B: Distribution of activity of the units for the image stimuli presented. The vertical lines represent the mean number of active neurons per image patch, their length the standard deviation.}
  \label{fig:Fig8}
\end{figure}

The degree of sparsity (and overcompleteness) is one of the determining factors in the differentiation of a homogeneous set of RFs into a more variable one with RFs with different functions~\cite{olshausen_learning_2009,olshausen_highly_2013}. Taking the two resulting classes obtained by CEL0 as a reference, we set the $\lambda$ parameter of ISTA to reproduce (approximately) the same distribution of activity as CEL0 (Fig~\ref{fig:Fig8}B). We refer to this choice as sparse ISTA.  We found that, unlike its original instantiation (Fig~\ref{fig:Fig7}C), sparse ISTA produced both circular and oriented RFs (Fig~\ref{fig:Fig7}B and \nameref{S7_Fig} for all the RFs), with a MSE approximately 3 times worse than the original instance of ISTA and of CEL0.

\subsubsection*{Impact of overcompleteness on reconstruction performance}
\begin{figure}[h!]
  \centering
      \includegraphics[width=\textwidth]{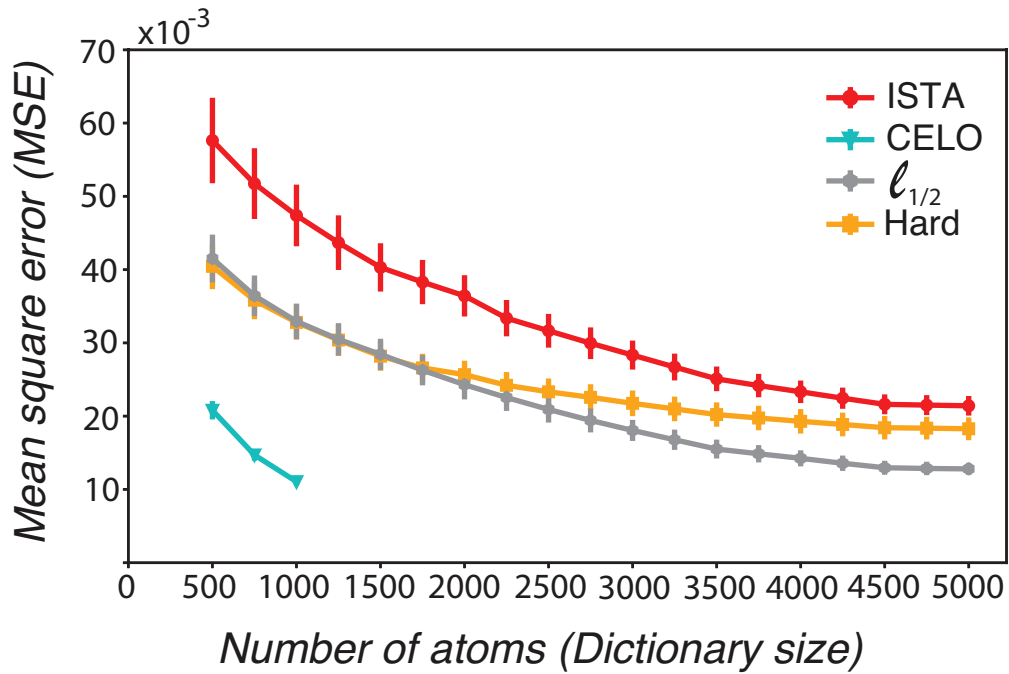}
  \caption{{\bf All methods have better reconstruction error than $\ell_1$-type reconstruction for all dictionary sizes tested.} The plot shows the MSE of the different thresholding methods for several dictionary sizes.  The parameter $\lambda$ has been adjusted for each dictionary size and algorithm so that the sparsity level is approximately stable (see \nameref{S8_Fig}). Each time we run $1600$ batches, with $250$ patches in each patch (in total $400000$ patches), and took the mean and standard deviation of the reconstruction error of the last $100$ batches.} 
  \label{fig:Fig9}
\end{figure}

As a final test, we asked how the reconstruction performance of the thresholding algorithms improved as we increased their degree of overcompleteness, i.e. the number of atoms while keeping the level of sparsity  constant (see \nameref{S8_Fig} for a more detailed explanation, and \nameref{tableParamsAll} for the $\lambda$ values for the different methods and number of units). We found that the $\ell_1$ model needs approximately $5000$ units to reach the reconstruction performance of CEL0 for $500$ units (Fig ~\ref{fig:Fig9}). We also observed that for dictionary sizes greater than $2000$ units (i.e., greater than $4\times$ degrees of overcompleteness) $\ell_{1/2}$ thresholding has a smaller reconstruction error than hard thresholding, with both performing better than ISTA for any dictionary size. CEL0 provides the best reconstruction up until $1000$ units. For more units at this level of sparsity, the algorithm did not converge for certain patches, and the results were not reported.

\section*{Discussion}

We show that continuous, non-convex thresholding-based algorithms produce much sparser activations for the same reconstruction error compared to the classically used soft thresholding algorithm ISTA making use of the convex $\ell_1$ regularizer. Among the penalties tested, CEL0 shows robustness to various levels of noise, producing sparse codes even for large levels of noise, with $\ell_{1/2}$ and hard thresholding following in terms of performance. All iterative algorithms maintain their sparse coding level independently on the dictionary used.

Sparsity is an optimization principle that pushes the RFs to diverge from a monolithic functionality into a richer variety~\cite{olshausen_learning_2009}, a characteristic that is present in V1~\cite{ringach_spatial_2002}. CEL0 produced this diversity for the smallest degree of overcompleteness ($\sim 2\times$) for a set reconstruction error. When the same sparsity level is enforced for ISTA, its reconstruction error is about $3$ times worse than CEL0. We then maintained the same sparsity for different dictionary sizes and found that soft thresholding performs worse in terms of reconstruction performance against all three competing methods. While CEL0 provides the best reconstruction performance, $\ell_{1/2}$ and hard thresholding converge to a sparse solution for a wider range of dictionary sizes.  

Natural image statistics show complex redundancy, and an underlying purpose of sparse neural codes is to get rid of in their representations~\cite{attneave_informational_1954, barlow_possible_1961,simoncelli_natural_2001}, by replacing it with a simpler form of redundancy, a prior on the individual neuron activity that is highly peaked at zero~\cite{olshausen_sparse_1997}. In non-convex regimes, the distributions of neuronal activities move from Laplace-type to Dirac-like~\cite{rehn_network_2007}. Our results suggest that the closer the units reach this distribution of activity, the better the generative model will match the diversity of RF shapes found in V1. It will be interesting to examine how this level of sparsity fits with neuron models that capture the highly nonlinear operations of dendrites~\cite{poirazi2020,bertalmio2020}

We see that sparsity produces secondary effects in V1, such as orientation selectivity and variability in the RFs of neurons. This appears to be a common strategy in the brain. For example, it is shown computationally that homeostatic processes, which aim to balance the activity in the brain, also generate neural networks that endow context sensitivity to RFs  ~\cite{rentzeperis_adaptive_2022}, and connectivity patterns with different degrees of specificity, flexibility, and robustness ~\cite{rentzeperis_adaptive_2021}.  

We finally highlight that Rozell and colleagues have showed in~\cite{rozell_sparse_2008} that the coding step solved by iterative thresholding algorithms can be better reformulated as a recurrent neural network, suggesting that an implementation producing sparse codes is plausible in the brain. In a neural network, indeed, a unit that is active inhibits other units with similar response features, a process that effectively minimizes the energy function \eqref{eqEnergy}. Compared to feedforward topologies, prevalent in deep neural networks, this kind of recurrent competition induces lateral inhibition between units and has been shown to be more robust against noisy stimuli and adversarial attacks~\cite{paiton_selectivity_2020}. Our results indicate that an inhibition stronger than the one induced by the $\ell_1$ penalty could push for the generation of more diverse RFs and is in agreement with recent experimental findings~\cite{yoshida_natural_2020}.

\newpage

\section*{Supporting information}

% Include only the SI item label in the paragraph heading. Use the \nameref{label} command to cite SI items in the text.

\begin{figure}[H]
  \centering
      \includegraphics[width=\textwidth]{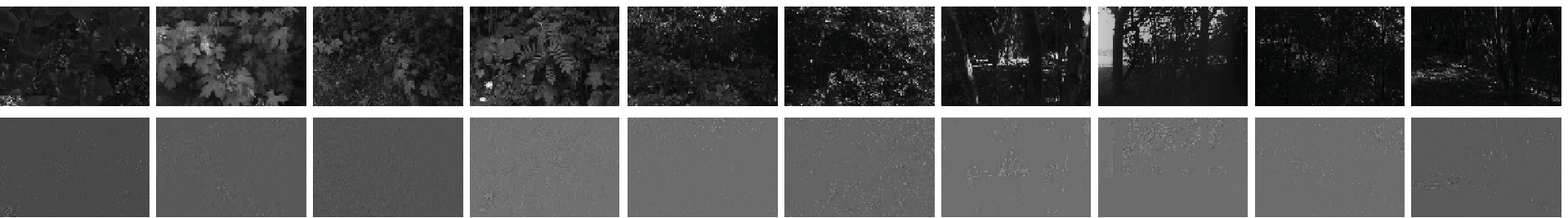}
\end{figure}

\paragraph*{S1 Fig.}
\label{S1_Fig}
{\bf Examples of raw and whitened images.} First row shows examples of raw images from the van Hateren database. Second row shows the corresponding whitened images used in the generative model.

    \begin{table}[H]
        \centering
        \scalebox{1.2}{
        \begin{tabular}{|c|c|c|c|c|}
        \hline
        \rowcolor{Gray}
        Methods & ISTA & $\ell_{1/2}$ & Hard & CEL0 \\
        \hline
        $\mu$ & $0.01$ &$0.01$ & $0.01$ & $0.1$\\
        \hline
        $\lambda$  & $0.41$ &$0.13$ & $0.013$ & $0.45$\\
        \hline
        \end{tabular}}

    \end{table}

\paragraph*{S1 Table}
\label{tableParams500}
{\bf Values of learning rate $r$, $\mu$ and relative weight $\lambda$ for the different algorithms for 500 units.}

\begin{figure}[H]
  \centering
      \includegraphics[width=\textwidth]{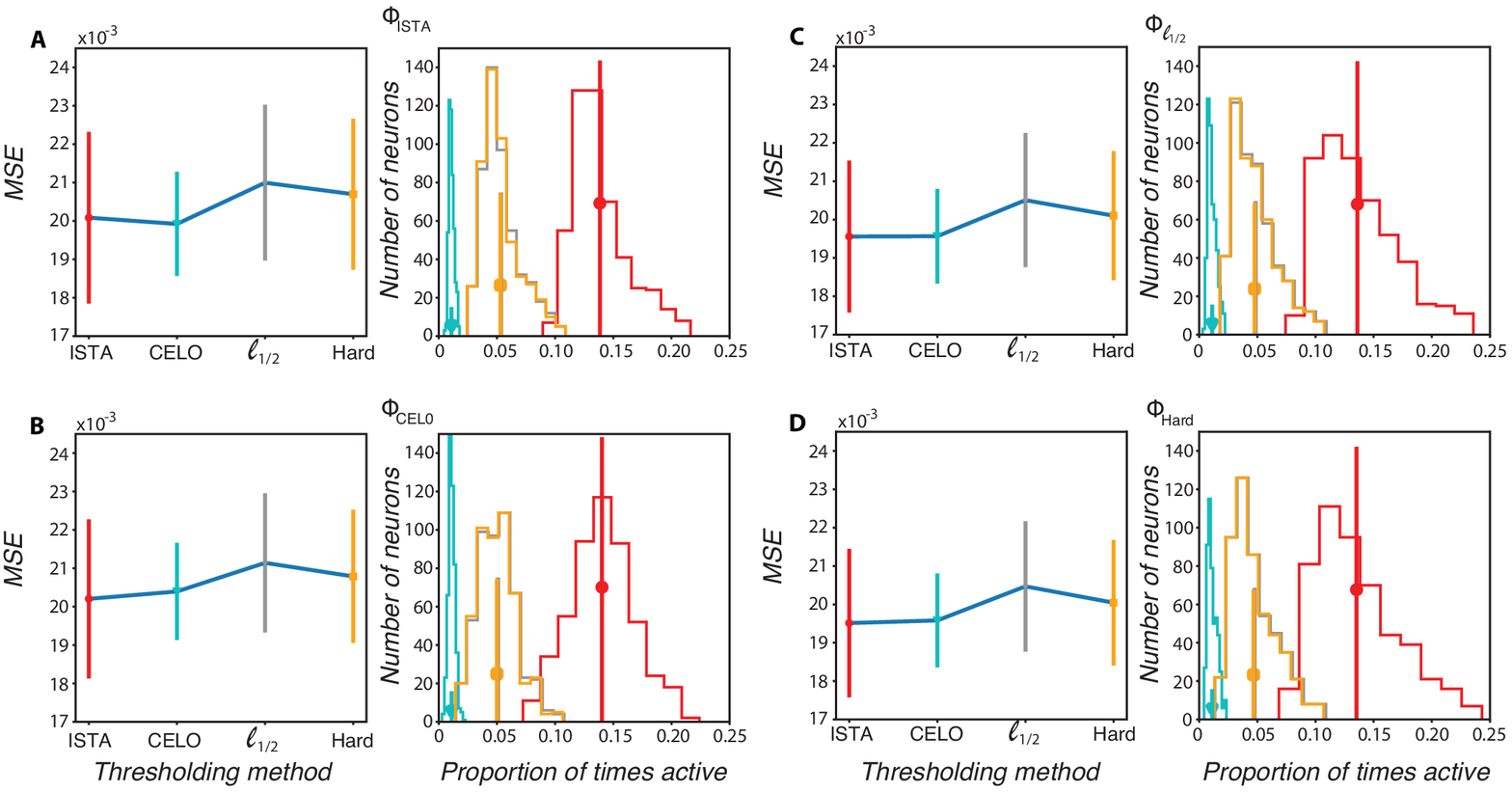}
\end{figure}

\paragraph*{S2 Fig.}
\label{S2_Fig}
{\bf Sparsity and reconstruction error are not affected by the $\Phi$ dictionary used for coding.} A: (left figure) MSE as a function of the thresholding method when $\Phi_{ISTA}$ is used as a dictionary. (right figure) Distribution of activity of the units for the image stimuli presented when $\Phi_{ISTA}$ is used as a dictionary. B: Same as A for $\Phi_{CEL0}$ as fixed dictionary C: Same as A for $\Phi{\ell_{1/2}}$ as fixed dictionary D: Same as A for $\Phi_{Hard}$ as fixed dictionary.

\begin{figure}[H]
  \centering
      \includegraphics[width=\textwidth]{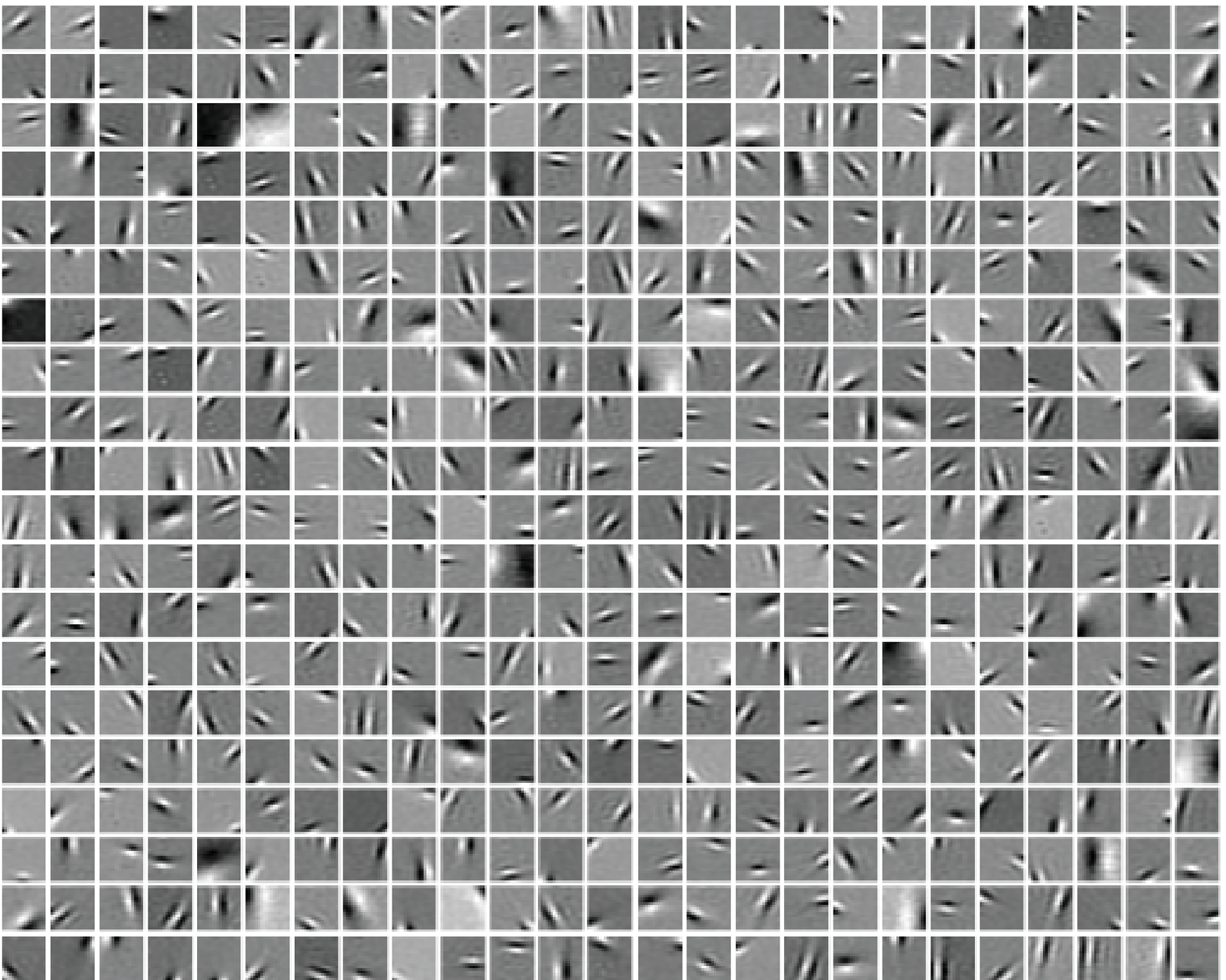}
\end{figure}

\paragraph*{S3 Fig.}
\label{S3_Fig}
{\bf 500 $\phi$ vectors learned from ISTA.}

\begin{figure}[H]
  \centering
      \includegraphics[width=\textwidth]{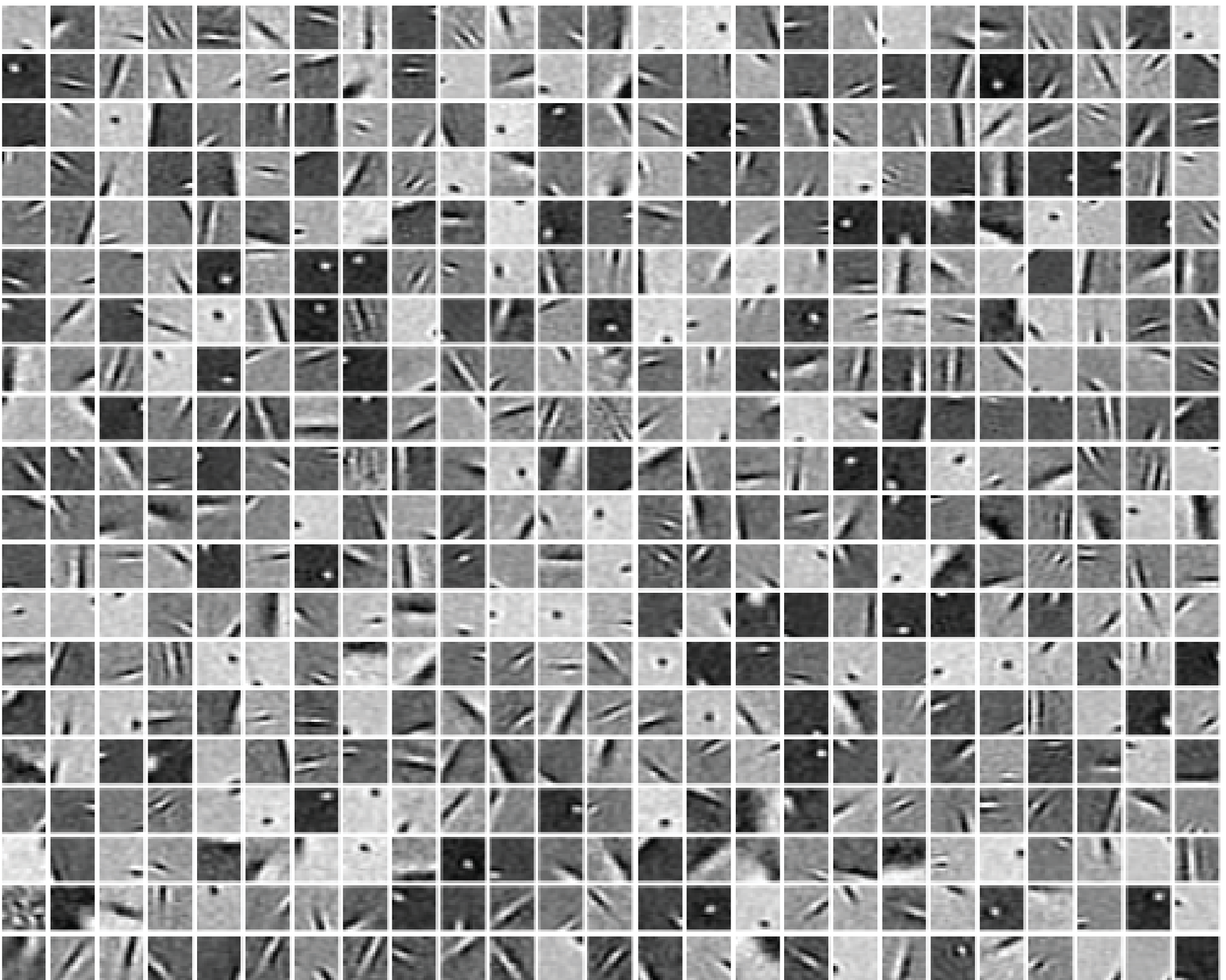}
\end{figure}

\paragraph*{S4 Fig.}
\label{S4_Fig}
{\bf 500 $\phi$ vectors learned from CEL0.}

\begin{figure}[H]
  \centering
      \includegraphics[width=\textwidth]{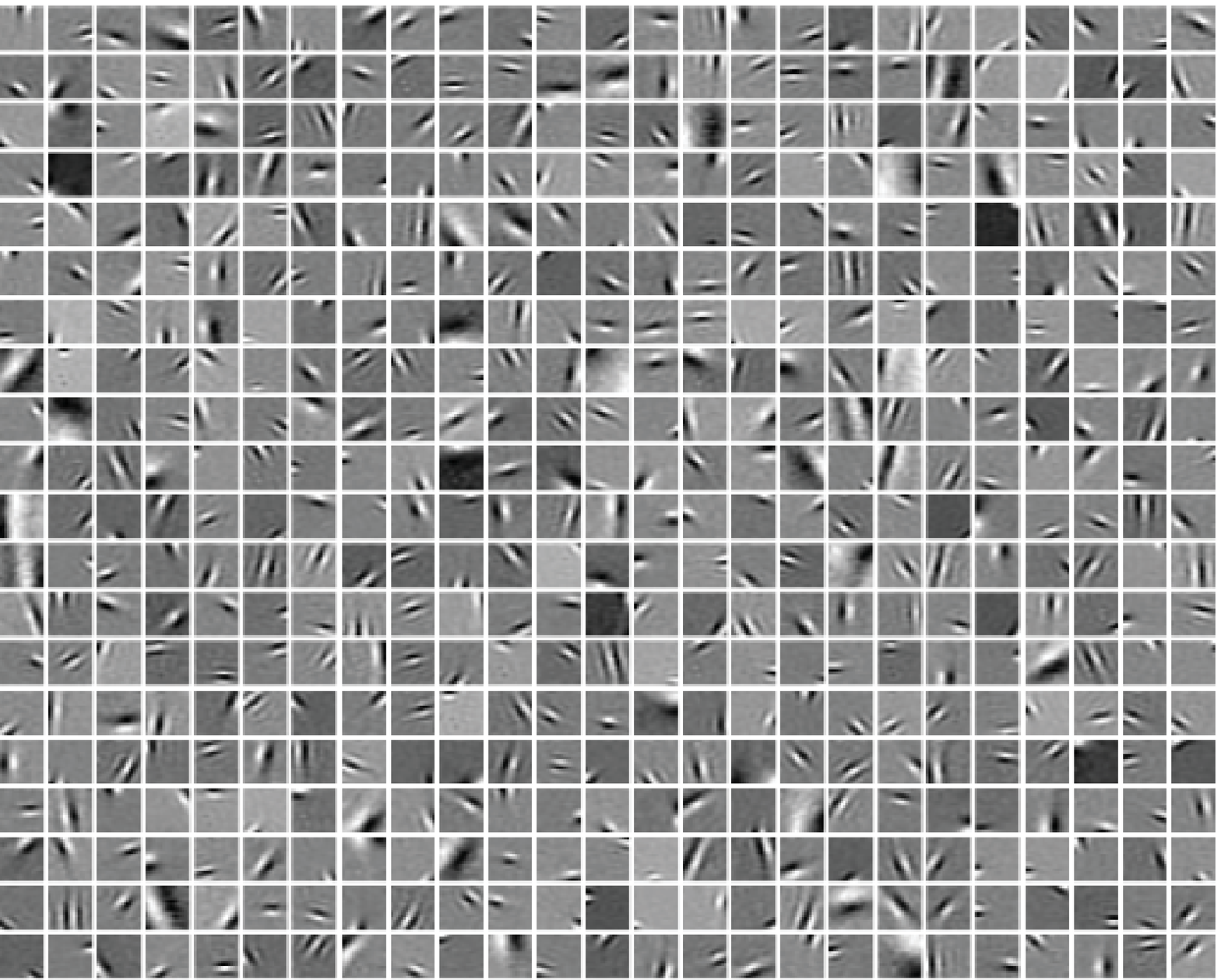}
\end{figure}

\paragraph*{S5 Fig.}
\label{S5_Fig}
{\bf 500 $\phi$ vectors learned from $\ell_{1/2}$ thresholding.}

\begin{figure}[H]
  \centering
      \includegraphics[width=\textwidth]{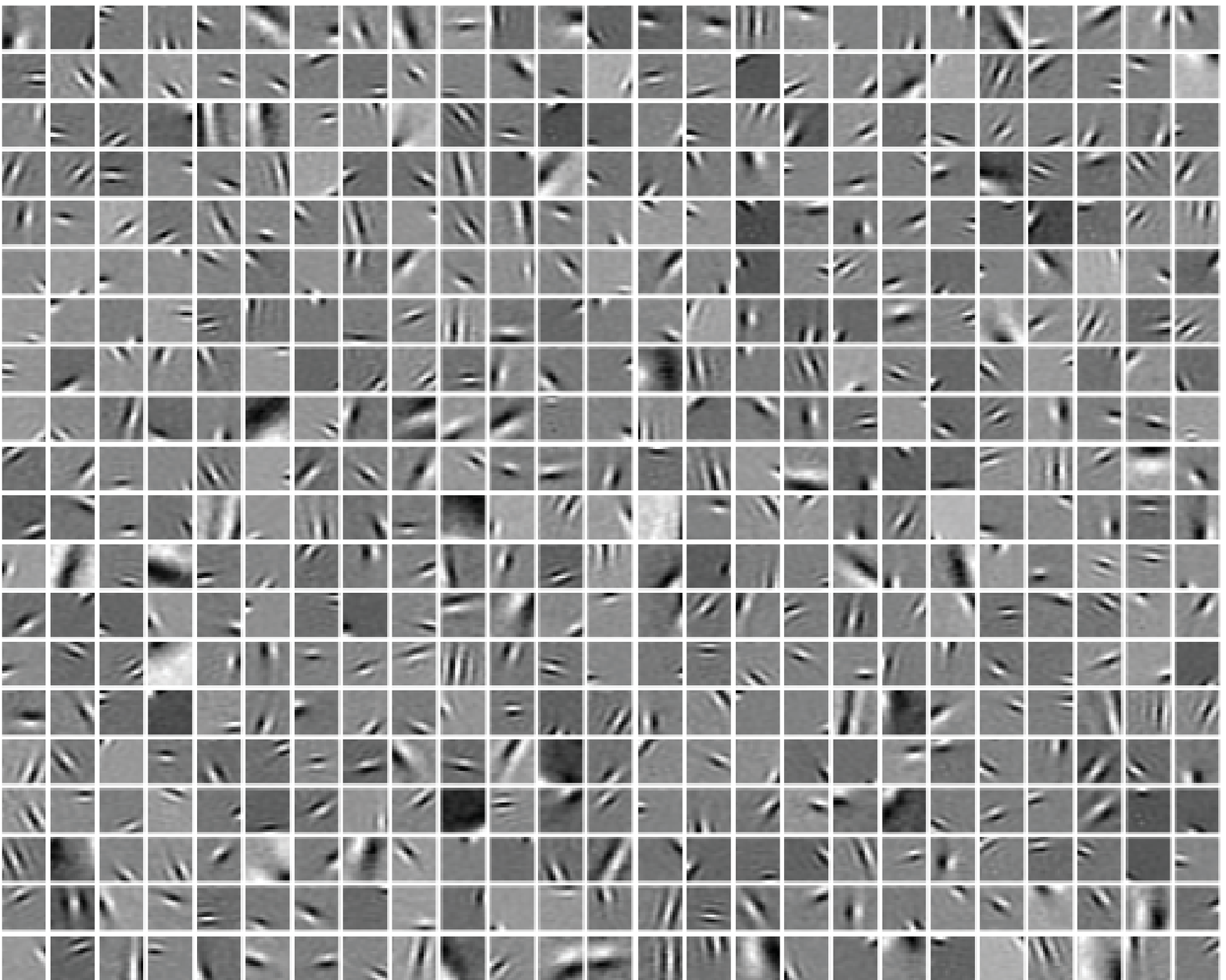}
\end{figure}

\paragraph*{S6 Fig.}
\label{S6_Fig}
{\bf 500 $\phi$ vectors learned from hard thresholding.}

\begin{figure}[H]
  \centering
      \includegraphics[width=\textwidth]{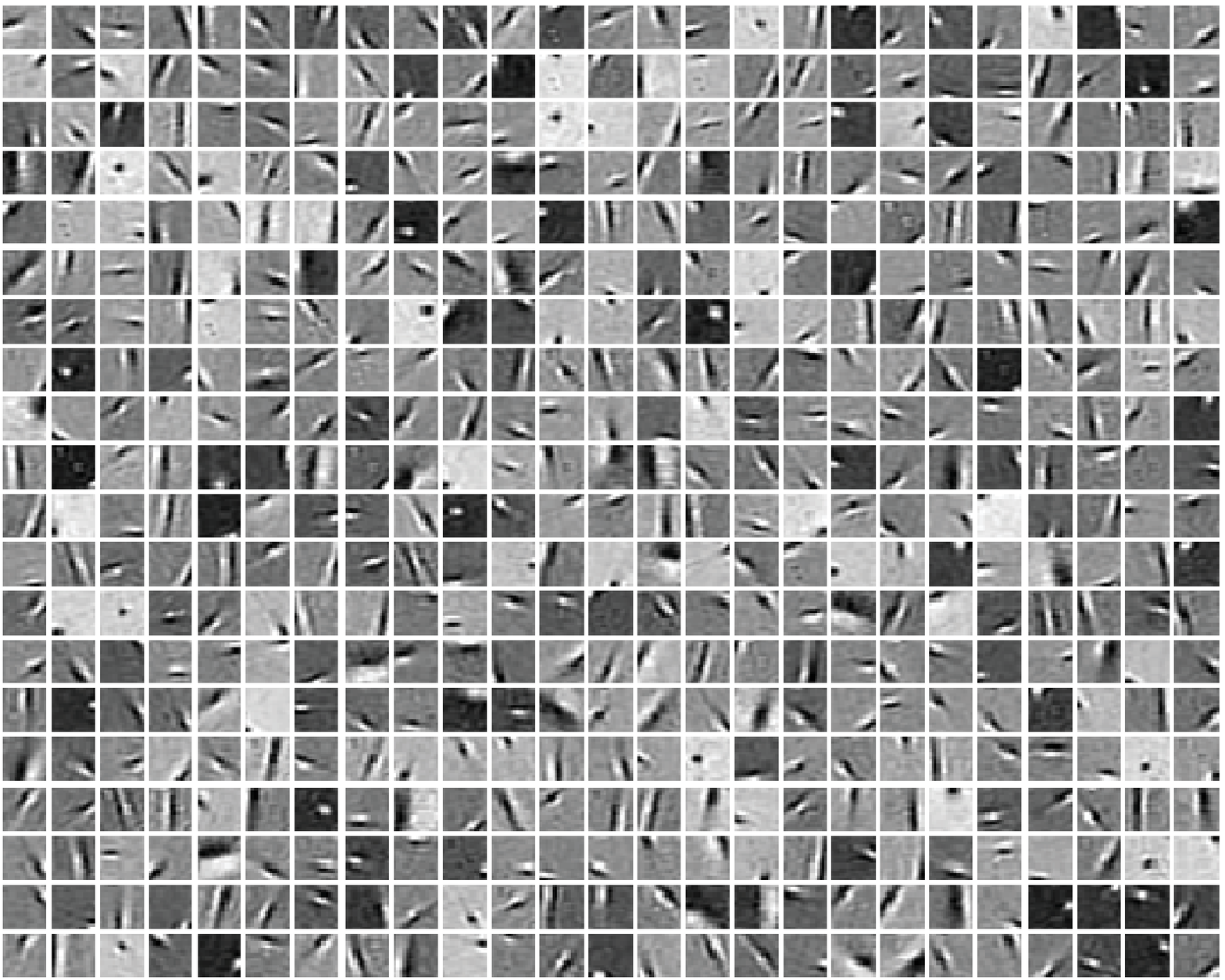}
\end{figure}

\paragraph*{S7 Fig.}
\label{S7_Fig}
{\bf 500 $\phi$ vectors learned from sparse ISTA thresholding.}

   \begin{table}[H]
        \centering
        \scalebox{1.2}{
        \begin{tabular}{|c|c|c|c|c|}
        \hline
        \rowcolor{Gray}
        Number of units & $\lambda_{ISTA}$ & $\lambda_{\ell_{1/2}}$ & $\lambda_{Hard}$ & $\lambda_{CEL0}$ \\
        \hline
         $500$ & $1.8$ &$0.35$ & $0.046$ & $0.45$\\
        \hline
        $750$  & $1.7$ &$0.33$ & $0.043$ & $0.44$\\
        \hline
         $1000$ & $1.6$ &$0.31$ & $0.04$ & $0.43$\\
        \hline
        $1250$  & $1.5$ &$0.295$ & $0.037$ & \\
        \hline
         $1500$ & $1.4$ &$0.285$ & $0.034$ & \\
        \hline
        $1750$  & $1.35$ &$0.275$ & $0.032$ & \\
        \hline
        $2000$ & $1.3$ &$0.265$ & $0.031$ & \\
        \hline
        $2250$  & $1.2$ &$0.255$ & $0.029$ & \\
        \hline
        $2500$ & $1.15$ &$0.245$ & $0.028$ & \\
        \hline
        $2750$  & $1.1$ &$0.235$ & $0.027$ & \\
        \hline
        $3000$ & $1.05$ &$0.225$ & $0.026$ & \\
        \hline
        $3250$  & $1$ &$0.215$ & $0.025$ & \\
        \hline
        $3500$ & $0.95$ &$0.205$ & $0.024$ & \\
        \hline
        $3750$  & $0.925$ &$0.2$ & $0.0235$ & \\
        \hline
        $4000$ & $0.9$ &$0.195$ & $0.023$ & \\
        \hline
        $4250$  & $0.875$ &$0.19$ & $0.0225$ & \\
        \hline
        $4500$  & $0.85$ &$0.185$ & $0.022$ & \\
        \hline
        $4750$ & $0.85$ &$0.185$ & $0.022$ & \\
        \hline
        $5000$  & $0.85$ &$0.185$ & $0.022$ & \\
        \hline 
        \end{tabular}}

    \end{table}

\paragraph*{S2 Table}
\label{tableParamsAll}
{\bf Values of the $\lambda$ parameter for the different algorithms and number of units.} The values for $\mu$, are the same for different number of units as shown in \nameref{tableParams500}.

\newpage
\begin{figure}[H]
  \centering
      \includegraphics[width=\textwidth]{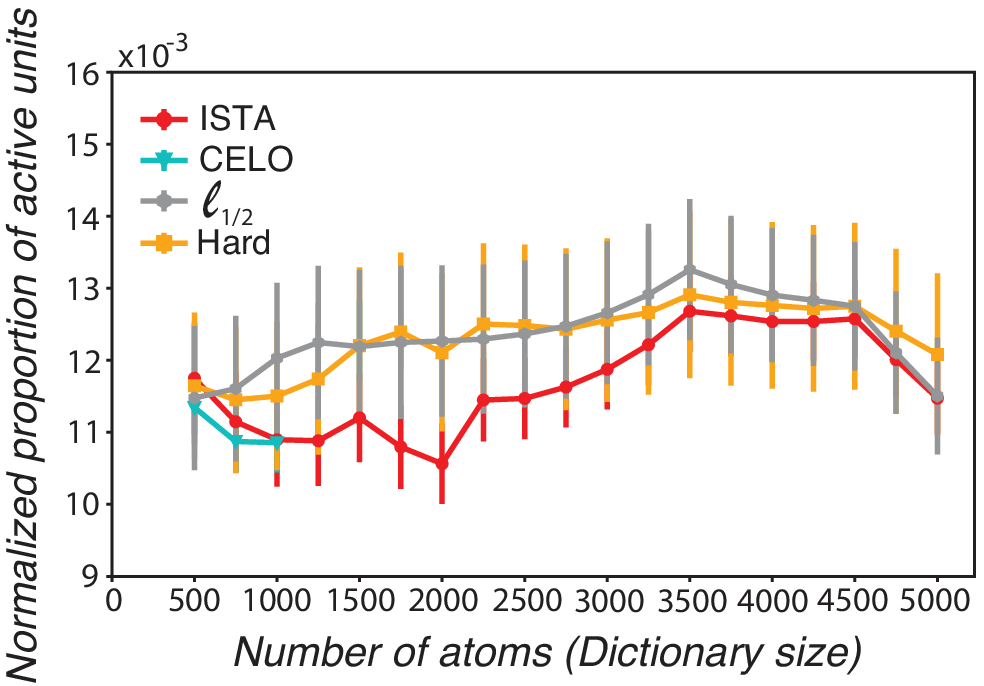}
\end{figure}

\paragraph*{S8 Fig.}
\label{S8_Fig}
{\bf Sparsity was constrained in a narrow window for all the dictionary sizes.} Normalized proportion of active units for the different number of atoms tested. The normalized proportion of active units is defined as the average number of active units for an image patch over the total number of units in the dictionary. The values taken by the parameter was between 0.0105 and 0.0132. Vertical indicates standard error from mean.

\section*{Acknowledgments}
All authors acknowledge the support received by the ANR JCJC grant RUBIN-VASE. LP received funding from the ANR grant number ANR-20-CE23-0021 ``{\sc AgileNeuroBot}''. LC acknowledges the support received by the GdR ISIS grant SPLIN.

\section*{Data Availability}
The code where one can run the generative model with the different thresholding functions is publicly available in \href{https://github.com/rentzi/sparseRegularizers}{https://github.com/rentzi/sparseRegularizers}.

%\end{thebibliography}

\end{document}